\documentclass[article, nojss]{jss}\usepackage[]{graphicx}\usepackage[]{color}
\makeatletter
\def\maxwidth{ %
  \ifdim\Gin@nat@width>\linewidth
    \linewidth
  \else
    \Gin@nat@width
  \fi
}
\makeatother

\usepackage{Sweave}

\usepackage{amsmath, bm, amssymb}
\usepackage{natbib}
\usepackage{booktabs}

\author{\begin{tabular}{*{2}{>{\centering}p{.5\textwidth}}}
\large Nome1 & \large Nome2 \tabularnewline
Department1 & Department2 \tabularnewline
School1 & School2 \tabularnewline
\url{url1} & \url{url2}
\end{tabular}
}

\author{
Yujing Jiang\\ University of Connecticut \And
  Xin He\\ University of Maryland \And
  Mei-Ling Ting Lee\\ University of Maryland \AND
  Bernard Rosner\\ Harvard University  \And
  Jun Yan\\ University of Connecticut
  }
\title{Wilcoxon Rank-Based Tests for Clustered Data with \proglang{R} Package
\pkg{clusrank}}

\Plainauthor{Yujing Jiang, Mei-Ling Ting Lee, Xin He,
  Bernard Rosner, Jun Yan}
\Plaintitle{Wilcoxon Rank-Based Tests for Clustered Data with R package clusrank}
\Shorttitle{Rank-Based Tests for Cluster Data} 

\Abstract{
Wilcoxon Rank-based tests are distribution-free alternatives
to the popular two-sample and paired t-tests.
For independent data, they are available in several \proglang{R}
packages such as \pkg{stats} and \pkg{coin}.
For clustered data, in spite of the recent methodological
developments, there did not exist an \proglang{R} package
that makes them available at one place.
We present a package \pkg{clusrank} where the latest developments are
implemented and wrapped under a unified user-friendly interface.
With different methods dispatched based on the inputs, this package
offers great flexibility in rank-based tests for various clustered data.
Exact tests based on permutations are also provided for some methods.
Details of the major schools of different methods are briefly reviewed.
Usages of the package \pkg{clusrank} are illustrated with simulated data
as well as a real dataset from an ophthalmological study.
The package also enables convenient comparison between
selected methods under settings that have not been studied
before and the results are discussed.
}

\Plainkeywords{Wilcoxon rank-sum test, Wilcoxon signed-rank test}

\Address{
  Yujing Jiang and Jun Yan\\
  Department of Statistics\\
  Professor of Statistics\\
  University of Connecticut\\
  215 Glenbrook Rd. Unit 4120, Storrs, CT 06269, USA.\\
  Email: \email{yujing.jiang@uconn.edu}, \email{jun.yan@uconn.edu}

  Xin He and Mei-Ling Ting Lee\\
  Department of Epidemiology and Biostatistics\\
  University of Maryland\\
  EPIB 2234R, SPH Building \#255,  College Park, MD 20742, USA.\\
  Email: \email{xinhe@umd.edu}, \email{mltlee@umd.edu}

  Bernard Rosner\\
  Department of Biostatistics\\
  Harvard University\\
  180 Longwood Avenue,  Boston, Massachusetts 02115, USA\\
  Email: \email{stbar@channing.harvard.edu}

}

\IfFileExists{upquote.sty}{\usepackage{upquote}}{}
\begin{document}


\section{Introduction}

The Wilcoxon rank-sum and signed-rank tests are important tools for
two-group comparisons and paired comparisons, respectively.
Unlike their counterparts under the normality assumption, they are attractive
because they are rank-based without the need of distributional assumptions.
Nonetheless, standard versions of such tests presume independent data, and
cannot be applied to clustered data which frequently arise in many fields.
Clustered data consist of data obtained from correlated observations
from sub-units or members in each cluster, where clusters may be
independent but measures from members within each cluster are not.
For example, in longitudinal studies or familial studies,
measures from observations of the same subject or
the same family are not independent but correlated.
The effective sample size for clustered data will be different from
the number of  observations in clusters due to intracluster dependence.
Often times, because of the positive intracluster dependence, the
variances of the test statistics are underestimated, and as a consequence,
the resulting p-values are smaller than what they should be.
The popular generalized estimating equations (GEE) approach
\citep{Lian:Zege:long:1986} provides a general regression
modeling strategy that considers intracluster dependence,
and can be used to compare two groups as a special case.
Unlike rank-based procedures, however, it is not invariant to
monotonic transformations of the data.

Several recent developments have extended the Wilcoxon rank-sum
test to allow two-sample comparisons for clustered data.
\citet{Rosn:Grov:use:1999} proposed a Mann--Whitney $U$-statistic for
clustered data which corrects the variance of the test statistic for four
types of intracluster correlation, but did not provide large sample theory.
\citet{Rosn:Glyn:Lee:inco:2003} proposed an extended Wilcoxon rank-sum
test under the assumptions that all sub-unit observations (or members)
from the same cluster (i.e., subject) belong to the same treatment group,
that observations within any cluster are exchangeable, and that the
intracluster dependence does not vary across groups.
With a test statistic in similar form as the standard Wilcoxon rank-sum
test after ranking all the observations combined, they derived the
asymptotic mean and variance under the clustered setting that
accommodates unequal cluster sizes and possible stratification.
\citet{Rosn:Glyn:Lee:exte:2006} extended their (2003) approach to
accommodate the situation where members of a single cluster may
be assigned to different treatment groups, but still assumes
exchangeability with the same intracluster dependence across groups.
The assumptions of \citet{Rosn:Glyn:Lee:inco:2003} were relaxed in the
approach of \citet{Datt:Satt:rank:2005}, which is based on
within-cluster resampling \citep{Hoff:Sen:Weib:with:2001} and remains
valid when the cluster sizes are informative.
More recently, \citet{Dutt:Datt:rank:2016} extended the idea of
within-cluster resampling to further accommodate the case where the
number of members in a group within a cluster is informative.

On the other hand, for the one-sample problems or paired comparison
problems, \citet{Rosn:Glyn:Lee:wilc:2006} extended the Wilcoxon
signed-rank test to clustered data by adjusting the variance of the standard
test statistic, assuming a common intracluster correlation across clusters.
The cluster sizes are allowed to vary, but the method does not
consider the informative cluster sizes where the distribution of
paired difference within a cluster depends on the cluster size.
\citet{Datt:Satt:sign:2008} proposed a signed-rank test based on sampling
members within cluster that accounts for informative cluster sizes.

Existing implementations of Wilcoxon rank-sum and signed-rank tests
are mostly standard versions where the data are assumed to be independent.
They have long been available in standard software, such as \code{wilcox.test}
in the built-in package \pkg{stats} of \proglang{R} \citep{R:2016},
\code{PROC NPAR1WAY} of \proglang{SAS} \citep{SAS:2013}, and
\code{ranksum} and \code{signrank} of \proglang{Stata} \citep{stata:2015}.
Permutation methods are available in \pkg{StatXact} \citep{Cytel:2013}.
These implementations cannot, however, handle clustered data in general.
One exception is, for instance, multi-center randomized clinical trials,
where the centers can be viewed as blocks across the treatment groups.
In this case, the \proglang{R} package \pkg{coin} \citep{Hoth:etal:impl:2008},
which provides a powerful toolkit for conditional inferences, can be applied.
The general situation where the sampling unit is cluster, however,
is not under the inference framework of the package \pkg{coin}.

Despite the popularity of clustered data arising from a wide range of
applications such as biomedical and social science studies,
the extensions of Wilcoxon rank-sum and signed-rank tests reviewed
above have not been implemented in \proglang{R} until very recently.
The package \pkg{clusrank} that we developed made first appearance
on the Comprehensive R Archive Network (CRAN) in December, 2015.
The package provides implementation of these recently available rank-sum
tests \citep{Rosn:Glyn:Lee:inco:2003, Datt:Satt:rank:2005,
  Rosn:Glyn:Lee:exte:2006} and signed-rank tests
\citep{Rosn:Glyn:Lee:wilc:2006, Datt:Satt:sign:2008} for clustered data.
The methods are grouped into two categories by their authors: RGL for
those by Rosner, Glynn, and Lee; and DS for those by Datta and Satten.
Note that the RGL methods are
available in \proglang{SAS} codes from Dr. Rosner's website
(\url{https://sites.google.com/a/channing.harvard.edu/bernardrosner/channing/})
and in \proglang{Stata} \citep{stata:2015} package \pkg{cluswilcox} from
Dr. Lee's website (\url{http://cls.umd.edu/mtlee/}).
\proglang{R} codes for the DS methods are available from Dr. Datta's website
(\url{http://www.somnathdatta.org/software/}), which was recently
put into \proglang{R} package \pkg{ClusterRankTest}
\citep{Dutt:Datt:clus:2016} in April, 2016.
Modeled after the familiar function \code{wilcox.test} in
the \proglang{R} built-in package \pkg{stats}, the package \pkg{clusrank}
that we developed unifies both RGL and DS methods under a user-friendly
interface that accommodates the specifications from these methods.
This makes it very easy for users to compare the performances of different
approaches under various settings; see Section~\ref{s:simu}.

The rest of the article is organized as follows.
The recently available Wilcoxon rank-sum tests and
signed-rank tests for clustered data are briefly reviewed in
Sections~\ref{s:ranksum} and~\ref{s:signedrank},  respectively.
The usage of the unified user-level function and major
input arguments are described in Section~\ref{s:usage}.
Illustrations of how to access the implemented methods
using both simulated data and a real dataset from an
ophthalmological study are presented in Section~\ref{s:illus}.
In Section~\ref{s:simu}, comparisons of selected methods that
have not been studied previously are reported, which
are made very easy by the package using only a few lines of codes.
A discussion concludes in Section~\ref{s:disc}.

\section{Rank-sum test for clustered data}
\label{s:ranksum}

The Wilcoxon rank-sum test is used for two-sample comparison.
Let $X_{ij}$ be the $j$th observation in the $i$th cluster,
$1 \leq i \le N$, $1 \le j \leq n_i$,
where $n_i$ is the size of cluster $i$.
Let $\delta_{ij}$ be the group indicator of $X_{ij}$;
$\delta_{ij} = 1$ if $X_{ij}$ is in group~1, and
$\delta_{ij} = 0$ if $X_{ij}$ is in group~2.
Let $R_{ij}$ be the rank of $X_{ij}$ among all the observations.
The observed data consist of
$(\mathbf{X}, \boldsymbol{\delta}) =
\{X_{ij}, \delta_{ij}: 1 \le j \le n_i; 1 \le i \le N\}$.
Clusters are assumed to be independent, while
subunit observations within each cluster are not.
The null hypothesis $H_0$ to be tested is that there is no difference
of the measures of location of the two groups; i.e., the distribution of
$X_{ij}$ remains the same regardless of the group indicator $\delta_{ij}$.

\subsection{RGL method with cluster-level grouping}
The RGL Wilcoxon rank-sum test \citep{Rosn:Glyn:Lee:inco:2003} was
designed for the scenario where the treatment group is assigned at the
cluster-level: i.e., $\delta_{ij} = \delta_i$ for all $1 \le j \le n_i$.
Define $R_{i+} =\sum_{j=1}^{n_i}R_{ij}$, the sum of observed ranks of
the subunits in the $i$th cluster.
The Wilcoxon rank-sum statistic is
\begin{equation}
\label{eq:rglrs}
W = \sum_{i = 1}^N \delta_{i}R_{i+}.
\end{equation}

The rationale of the RGL test procedure is random permutation
conditioning on the observed $R_{i+}$, $i = 1, \ldots, N$.
Like all permutation-based approaches, the RGL method assumes that
subunit observations within each cluster are exchangeable and that the
intracluster dependence remains the same across groups.
To derive the sampling distribution of $W$ given $R_{i+}$'s,
\citet{Rosn:Glyn:Lee:inco:2003} stratified on the cluster sizes and
investigated $R_{i+}$ between two groups for each cluster size.
Let $G$ be the maximum cluster size; i.e., $G = \max_{1 \le i \le N} n_i$.
Then $W$ in Equation~\eqref{eq:rglrs} can be written as
\begin{equation}
\label{eq:rglrs1}
W = \sum_{g=1}^{G}\sum_{i \in I_g}\delta_{i}R_{i+} = \sum_{j = 1}^{G}W_g,
\end{equation}
where $I_g$ is the set of indices of clusters whose size is $g$ and
$W_g = \sum_{i \in I_g} \delta_i R_{i+}$.
The null distribution of $W$ conditioning on $R_{i+}$'s is
obtained by combining all possible permutations of $W_g$ for
each cluster size $g \in \{1, \ldots, G\}$.
Let $N_g$ be the number of clusters of size $g$,
among which $m_g$ are in group~1 and $n_g$ are in group~2.
The total number of permutation is
$\prod_{g = 1}^G {N_g \choose m_g}$.

When $N$ is large, exhaustive permutation is infeasible, and
\citet{Rosn:Glyn:Lee:inco:2003} proposed an asymptotic test statistic
$Z = {(W - \E (W)) } /  {\sqrt{\VAR(W)}},$
where $\E(W) = \sum_{g = 1}^G \E(W_g)$, and
$\VAR(W) = \sum_{g=1}^G \VAR(W_g)$.
Under $H_0$, for clusters with size $g$, the distribution of
$\delta_i$ is Bernoulli with probability $m_g/N_g$, and we have
$\E(\delta_i) = m_g / N_g$,
$\VAR(\delta_i) = m_g n_g / N_g^2$, and
$\COV(\delta_i, \delta_j) = - m_g n_g / [N_g^2(N_g - 1)]$.
The specific expressions needed are shown to be
$\E(W_g) = m_g R_{++, g}  / N_g$ and
\[
\VAR(W_g) = \frac{m_g n_g }{N_g(N_g - 1)}
\sum_{i \in I_g} \left(R_{i+} - \frac{R_{++,g}}{N_g}\right)^2,
\]
where $R_{++,g} = \sum_{i\in I_g} R_{i+}$.
\citet{Rosn:Glyn:Lee:inco:2003} showed that under mild conditions, the
asymptotic distribution of the test statistic $Z$ is standard normal.
The method can be extended to the case of stratified data.

\citet{Rosn:Glyn:Lee:inco:2003} compares groups at each cluster size, the
imbalance of sample sizes between two groups across cluster size
strata may result in inefficiency \citep[p.911]{Datt:Satt:rank:2005}.
If only one group shows up at a cluster size, the corresponding data
will be ignored as no comparison at this cluster size can be made.
Further, the rank-sum statistic scores clusters by the sum of ranks of
cluster members, which is expected to perform best if intracluster
dependence is weak; otherwise, when the effective number of
independent observations per cluster becomes smaller, it overweights
larger clusters, and, hence, may have lower efficiency.

\subsection{DS method with subunit-level grouping}

Unlike the RGL method, the DS method allows subunit observations
within the same cluster to have different group memberships.
The rationale is rooted in the within-cluster resampling principal
of \citet{Hoff:Sen:Weib:with:2001}.
The test statistic is constructed from randomly picking one observation
from each cluster to form a pseudo-sample and averaging the standard
Wilcoxon rank-sum statistic over all possible pseudo-samples.
Let $X_i^*$ be a random pick from the $i$th cluster in the
pseudo-sample, and $\delta_i^*$ be its group membership.
The Wilcoxon rank-sum statistic for the pseudo-sample is
\begin{equation*}
  W^* = \frac{1}{N+1}\sum^N_{i = 1}\delta_i^* R_i^*,
\end{equation*}
where $R_i^*$ is the rank of $X_i^*$ among the pseudo-sample.
The test statistic of the DS method is
\begin{equation*}
  Z = \frac{S - \E(S)}{\sqrt{\widehat{\VAR}(S)}},
\end{equation*}
where $S = \E(W^* | \mathbf{X}, \boldsymbol{\delta})$.

\citet{Datt:Satt:rank:2005} derived the quantities needed to calculate
the test statistic, using mid-ranks to allow for ties in the data.
Let $F_i(x) = n_i^{-1} \sum_{k = 1}^{n_i} I(X_{ik} \le x)$ be the
empirical distribution function of the observations from cluster~$i$
and define $F_i(x-) = n_i^{-1} \sum_{k = 1}^{n_i} I(X_{ik} < x)$.
It can be shown that
\[
S = \frac{1}{N+1} \sum_{i=1}^N\sum_{k=1}^{n_i} \frac{\delta_{ik}}{n_i}
\left[1 + \frac{1}{2}\sum_{j\ne i}\Big\{F_j(X_{ik}) + F_j(X_{ik}-)\Big\}\right].
\]
The expectation turns out to be
\[
\E(S) = \E(W^*) = \frac{1}{2}\sum_{i=1}^N \frac{n_i} {N}.
\]
The variance term $\VAR(S)$ can be estimated by
$\widehat\VAR(S) = \sum_{i=1}^N\big\{ \hat W_i - \E(W_i)\big\}^2,$
where
\[
\hat W_i = \frac{1}{2 n_i (N + 1)}
\sum_{k=1}^{n_i}\left[(N - 1) \delta_{ik} - \sum_{j\ne i}^N \frac{n_{j1}}{n_j}\right]
\left[\hat F(X_{ik}) + \hat F(X_{ik}-)\right],
\]
\[
E(W_i) = \frac{N}{N + 1}\left(\frac{n_{i1}}{n_i} - \frac{1}{N}\sum_{j=1}\frac{n_{j1}}{n_j}\right),
\]
with $n_{j1}$ being the number of subunits in group~1 from cluster~$j$,
and $\hat F = \sum_{i=1}^N n_i F_i / \sum_{i=1}^N n_i$,
the pooled empirical distribution function of the observations.
The asymptotic distribution of $Z$ is standard normal under mild
conditions \citep[p.910]{Datt:Satt:rank:2005}.

The DS method can be generalized to the comparison of location among
$m$ treatment groups, $m \geq 3$, with the test statistic constructed
from a quadratic form of group-wise rank-sum vector.
This method allows arbitrary intracluster dependence structure (not
necessarily exchangeable as assumed in the RGL method) within each
cluster and remains valid when treatment affects the correlation structure.
However, this test cannot be applied to strictly contralateral data (e.g., when
each subject in an eye study have exactly one eye under each treatment)
due to violation of the assumptions required by the asymptotic theory.

\subsection{RGL method with subunit-level grouping}

\citet{Rosn:Glyn:Lee:exte:2006} extended the RGL method to allow
subunit-level grouping; i.e., for each cluster $i$,
treatment group indicator $\delta_{ij}$ may take different values for
$j = 1, \ldots, n_i$.
The idea of this method can be easily explained with balanced data,
where all cluster sizes are all the same; i.e., $n_i = g$ for all $i$.
The rank-sum statistic is
\begin{equation*}
  W_{g,N} = \sum_{i = 1}^N\sum_{j=1}^g\delta_{ij}R_{ij},
\end{equation*}
where $R_{ij}$ is the rank of $X_{ij}$ among all the observed data.
A cluster may have $q \in \{0, 1, \ldots, g\}$ subunits in group~1.
The sampling distribution of $W_{g,N}$ is derived from a two-stage
randomization: first, each cluster $i$ is randomly assigned to a
random number $Q_i$ according to the observed grouping distribution
$Q$; then, within cluster $i$, a random $Q_i$ out of $g$ subunits
are assigned to group~1 while all the rest are assigned to group~2.
Essentially, the first stage determines how many subunits are in
group~1 in a cluster and the second stage determines which they are.
The two-stage randomization process can be used to devise a random
permutation test to exhaust all possibilities for small $N$ and $g$.

It can be shown that
\[
\E(W_{g,N}) = \frac{gN + 1}{2} \sum_{q=0}^g q N_q,
\]
where $N_q$ is the number of clusters with $q$ members in group~1.
The variance of $W_{g,N}$ can be estimated by
\[
\widehat\VAR(W_{g,N}) = \frac{N^2}{(N - 1)g^2} \widehat\VAR(Q) s_B^2
+ N \hat\E[Q (g - Q)] s_W^2 / g,
\]
where
$s_B^2 = \sum_{i=1}^N \{R_{i+} - g(gN + 1) / 2\}^2 / N$,
$s_W^2 = \sum_{i=1}^N\sum_{j = 1}^g (R_{ij} - R_{i+} / g)^2 / \{N (g - 1)\}$,
and $\widehat\VAR$ and $\hat\E$ are operated on the empirical
distribution of $Q$.
\citet{Rosn:Glyn:Lee:exte:2006} showed that
\[
Z_{g,N} = \frac{W_{g,N} - E(W_{g,N})}{\sqrt{\widehat{\VAR}(W_{g,N})}}
\]
converges to a standard normal distribution $N \to \infty$
provided that $\lim_{N\to\infty} N_q / N = \xi_q$, where
$0 \le \xi_q \le 1$, if $0 < q < g$; or, $0 \le \xi_q < 1$, if
$q \in \{0, g\}$.
This test is equivalent to the RGL test for balanced data when the
treatment is assigned at the cluster-level.

For unbalanced data, let $N^{(g)}$ be the number of clusters of size
$g$ such that $N = \sum_{g=1}^G N^{(g)}$, where
$G = \max_{1\le i \le  N} n_i$ is the maximum cluster size.
A test procedure can be constructed by efficiently combining
$W_{g, N^{(g)}}$ across all $g \in \{1, \ldots, G\}$.
\citet{Rosn:Glyn:Lee:exte:2006} proposed to base the test on a
combined estimator $\hat\theta_N$ of $\theta$, the probability that an
observation in group~1 is greater than that of an observation in
group~2, which is 1/2 under the null hypothesis.
Standardized by a variance estimator $\widehat\VAR(\hat\theta_N)$,
the test statistic
$(\hat\theta_N- 1/2) / \widehat\VAR^{1/2}(\hat\theta_N)$
converges to a standard normal distribution as $N \to \infty$
under mild conditions.

\section{Signed-rank test for clustered data}
\label{s:signedrank}

The Wilcoxon signed-rank test is often used for paired data comparisons.
Let $X_{ij}$ be the paired-difference score for the $j$th pair in the
$i$th cluster,  $i = 1, \ldots, N$, $j = 1, \ldots, n_i$.
The null hypothesis $H_0$ is that the marginal
distribution of $X_{ij}$ is symmetric around $0$.
Note that unlike the rank-sum test, all subjects belong to a single
group in the paired comparison setting.
Let $R_{ij}$ be the rank of $|X_{ij}|$ among
$\{|X_{ij}|, i = 1, \ldots, N, j = 1 \ldots, n_i\}$.
Let $S_{ij} = V_{ij}R_{ij}$ be the signed-rank, where
$V_{ij} = \mathrm{sign}(X_{ij})$.

\subsection{RGL method: uninformative cluster size}
The RGL method for the rank-sum test \citep{Rosn:Glyn:Lee:inco:2003}
was adapted to the signed-rank test in \citet{Rosn:Glyn:Lee:wilc:2006}.
For balanced data where $n_i = g$ for all $i$,
the clustered WIlcoxon signed-rank statistic is
\[
T = \sum_{i=1}^N S_{i+} = \sum_{i=1}^N  \sum_{j=1}^g R_{ij} V_{ij},
\]
where $S_{i+} = \sum^g_{j=1}S_{ij}$ is the rank sum within the $i$th
cluster and only nonzero $X_{ij}$ are considered in the computation of
signed-ranks.
The null sampling distribution of $T$ can be obtained from a
randomization at the cluster-level conditional on $S_{i+}$'s.
Let $\delta_i$, $i = 1, \ldots, N$, be independent and identically
distributed random variables with equal probability being 1 and $-1$;
this distribution has variance 1.
The conditional distribution of $T$ given $S_{i+}$'s is the
same as that of $T_p = \sum_{i=1}^N \delta_i S_{i+}$.
For small $N$, it is possible to assess the significance of $T$ by
enumerating all $2^N$ possibilities and computing the tail probability.
A large sample test can be constructed with
$\E(T) = 0$ and $\VAR(T) = \sum_{i=1}^n S_{i+}^2$.
As $N \to \infty$, $T / \VAR^{1/2}(T)$ converges to a standard
normal distribution provided $g < \infty$.

For unbalanced data, \citet{Rosn:Glyn:Lee:wilc:2006} considered a
stratified statistic
\[
T = \sum_{i=1}^N w_i\bar{S}_i,
\]
where $\bar{S}_i = S_{i+} / n_i$ and $w_i = 1/\VAR(\bar{S}_i)$ under $H_0$.
The variance estimator $\widehat\VAR(\bar S_i)$ is obtained
assuming a shared intracluster correlation coefficient.
The randomization distribution of $T$ given $\bar{S}_i$'s is
that of
$T_p = \sum_{i=1}^N \delta_i w_i \bar{S}_i$,
which facilitates a random permutation test for small $N$.
The large sample test statistic is
$T / (\sum_{i=1}^N \hat w_i^2 \bar S_i^2)^{1/2}$,
where $\hat w_i = 1 / \widehat\VAR(\bar S_i)$.
It converges to a standard normal distribution as $N\to \infty$
provided that $G < \infty$ and $\lim_{N\to\infty} N_g / N \to \xi_g$
where $0 \le \xi_g \le 1$ for all $g \in \{1, \ldots, G\}$.
This method assumes that the cluster size distribution is
uninformative, and, hence, is not valid when the distribution of
paired differences within a cluster depends on the cluster size.

\subsection{DS method: informative cluster size}
\citet{Datt:Satt:sign:2008} followed the same principle of
within-cluster resampling as in \citet{Datt:Satt:rank:2005} in the
context of clustered paired data to develop a signed-rank test.
This method allows informative cluster sizes as long as the marginal
distributions of $X_{ij}$'s are identical for all $i$ and $j$.
Suppose that from the $i$th cluster, paired difference $X_{ij}$, denoted by
$X_i^*$, is randomly picked and pooled to form a pseudo-sample.
Let $R_i^*$ be the mid-rank of $|X_i^*|$, $i = 1, \ldots, N$ to allow
ties in the data, and let $V_i^* = \mathrm{sign}(X_i^*)$.
A standard Wilcoxon signed-rank test statistic for the pseudo-sample is
$\sum_{i=1}^N S_i^*$, where $S_i = V_i^* R_i^*$.
The DS method is based on
$T = \E (\sum_{i=1}^N S_i^* | \mathbf{X})$,
where $\mathbf{X} = \{X_{ij}: 1 \le i \le N; 1 \le j \le n_i\}$.

To compute $T$, let
\[
\hat H_i(x) = \frac{1}{2}\{ F_i(x) + F_i(x-)\},
\]
where $F_i(x)$ is the empirical distribution function of the observations
in cluster~$i$ at $x$ and $F_i(x-)$ is the left limit of $F_i(x)$ at $x$ as
in the DS method for the rank-sum test.
It turns out that
\[
T = \sum_{i=1}^N \frac{n_i^+ - n_i^-}{n_i}
+ \sum_{i=1}^N \frac{1}{n_i} \sum_{k=1}^{n_i} V_{ik} \sum_{j \ne i} H_j(|X_{ij}|),
\]
where $n_i^+ = \sum_{j=1}^{n_i} I(X_{ij} > 0)$ and
$n_i^- = \sum_{j=1}^{n_i} I(X_{ij} < 0)$.
The variance can be estimated by
$\widehat\VAR(T) = \sum_{i=1}^N \hat S_i^2$, where
\[
\hat S_i = \frac{n_i^+ - n_i^-}{n_i} +
\frac{N - 1}{n_i} \sum_{k=1}^{n_i} V_{ik} \hat H(|X_{ik}|),
\]
with $\hat H(x) = \sum_{i=1}^N n_i \hat H_i(x) / \sum_{i=1}^N n_i$.
The standardized test statistic
$Z = T / \sqrt{\widehat\VAR(T)}$ converges to a standard normal
distribution under mild conditions.

Note that the null distribution being tested using the DS method
is the distribution of the paired difference of a randomly selected
pair from a randomly selected cluster, regardless of the cluster size.
Whereas the null distribution of most signed-rank tests is the common
distribution of a randomly selected paired difference conditional on
the size of the cluster it belongs to.
Therefore the latter framework is a special case of the former.
Also, the DS method accounts for the cluster size by assigning equal
weight to each cluster instead of each paired difference, e.g., a paired
difference from a larger cluster will be assigned a smaller weight
than a paired difference from a smaller cluster.

\section{Usage}
\label{s:usage}

Package \pkg{clusrank} provides a unified interface to all the
methods reviewed in Sections~\ref{s:ranksum} and~\ref{s:signedrank}
through function \code{clusWilcox.test}:
\begin{Schunk}
\begin{Sinput}
R> library(clusrank)
R> args(clusWilcox.test)
\end{Sinput}
\begin{Soutput}
function (x, ...) 
NULL
\end{Soutput}
\end{Schunk}
Argument \code{x} can be either a numeric vector or a formula.
The default interface is called if \code{x} is a numeric vector, in
which case the interface is designed to mimic that of the function
\code{wilcox.test}:
\begin{Schunk}
\begin{Sinput}
R> args(getS3method("clusWilcox.test", "default"))
\end{Sinput}
\begin{Soutput}
function (x, y = NULL, cluster = NULL, group = NULL, stratum = NULL, 
    data = NULL, alternative = c("two.sided", "less", "greater"), 
    mu = 0, paired = FALSE, exact = FALSE, B = 2000, method = c("rgl", 
        "ds"), ...) 
NULL
\end{Soutput}
\end{Schunk}

The arguments \code{x}, \code{y}, \code{alternative}, \code{mu},
\code{paired}, and \code{exact} have the same meaning as those in the
familiar default interface of \code{wilcox.test}.
Clustered rank-sum test is requested if \code{paired = FALSE};
clustered signed-rank test is requested if \code{paired = TRUE}.
For both tests, the RGL and DS methods are requested with
\code{method} set to be \code{rgl} and \code{ds}, respectively.

For clustered rank-sum tests, \code{x}, \code{cluster}, and \code{group} are
required,  which are of the same length; \code{cluster} and \code{group}
specify the cluster membership and group membership, respectively.
Argument \code{y} is not used for clustered rank-sum tests.
The group assignment can be at either the cluster or the subunit level.
When using RGL method for data with treatment group assigned at the
cluster-level, an optional argument \code{stratum} can be specified to
account for the stratification and therefore provide a more powerful test.
The variables \code{x}, \code{cluster}, \code{group} and \code{stratum} can
be found from a data frame specified by argument \code{data}.

For clustered signed-rank tests, \code{x} and \code{cluster} are required
while \code{group} is not needed because the data are paird differences.
Argument \code{x} can be the parid difference between the pre- and
post-treatment observations; alternatively, \code{x} and \code{y} can
specify the pre- and post-treatment observations, respectively.
This interface is also similar to that of \code{wilcox.test}.

The exact version of the test is requested by \code{exact = TRUE}.
In this case, argument \code{B} is the number of random permutations to
approximate the exact test, with default value 2000.
The truly exact test is available for the RGL clustered rank-sum test with
cluster-level grouping and the RGL clustered signed-rank test, which can be
requested by setting \code{B = 0}; since this test is very computing
intensive, it is recommended to be used only with small samples.

The formula interface mimics that of the \code{wilcox.test} too
with arguments \code{formula}, \code{subset} and \code{na.action}:

\begin{Schunk}
\begin{Sinput}
R> args(getS3method("clusWilcox.test", "formula"))
\end{Sinput}
\begin{Soutput}
function (formula, data = parent.frame(), subset = NULL, na.action = na.omit, 
    alternative = c("two.sided", "less", "greater"), mu = 0, 
    paired = FALSE, exact = FALSE, B = 2000, method = c("rgl", 
        "ds"), ...) 
NULL
\end{Soutput}
\end{Schunk}
For clustered rank-sum tests, the left hand side of \code{formula} should be
the data vector to be tested, and the right hand side of \code{formula}
should contain the variables indicating group, cluster and possibly stratum.
Except for the group variable, other variables on the right need to be
indicated with special terms; for example, in formula
\code{z ~ group + cluster(cid) + stratum(sid)},
\code{group} identifies the grouping variable, \code{cid} identifies
the clusters, and \code{sid} identifies the stratum.
See Section~\ref{s:illus} for detailed illustrations.
For clustered signed-rank test, the left hand side of the formula is
the paired difference and the right hand side comprises the cluster id.
Neither \code{group} or \code{stratum} is applicable for signed-rank tests.
Other arguments are identical to those in the default interface.

\section{Illustrations}
\label{s:illus}

\subsection{Rank-sum test for clustered data}

We use a scheme that is similar to the simulation study in
\citet{Datt:Satt:rank:2005} to generate data for clustered
rank-sum test with both balanced and unbalanced data.
For group $g \in \{0, 1\}$, the observations in a cluster
is generated as
\[
X = \exp(Z_g) + \delta g,
\]
where $Z_g$ is a standard multivariate normal random vector with mean zero
and an exchangeable or autoregressive of order~1 (AR1) correlation matrix
with correlation parameter $\rho_g$, and $\delta$ is the group difference.
The AR1 correlation structure is common in longitudinal studies, and it can
be used to investigate the robustness of the methods when the intra-cluster
correlation is not exchangeable. A correlation matrix of dimension
\code{dim} with correlation parameter \code{rho} can be generated as
follows, with \code{ex} for exchangeable and \code{ar1} for AR1.
\begin{Schunk}
\begin{Sinput}
R> ex  <- function(dim, rho) {
+    diag(1 - rho, dim) + matrix(rho, dim, dim)
+  }
R> ar1 <- function(dim, rho) {
+    rho ^ outer(1:dim, 1:dim, function(x, y) abs(x - y))
+  }
\end{Sinput}
\end{Schunk}

Below is a simple implementation that allows different levels of grouping
(cluster-level and subunit-level) and unequal cluster sizes.
Package \pkg{mvtnorm} \citep{Genz:etal:mvtn:2016, Genz:Bret:comp:2009}
is used to generate multivariate normal random vectors.
\begin{Schunk}
\begin{Sinput}
R> library(mvtnorm)
R> datgen.sum <- function(nclus, maxclsize, delta = 0., rho = c(0.1, 0.1),
+    corr = ex, misrate = 0., clusgrp = TRUE) {
+      nn <- nclus * maxclsize
+      Sigma1 <- corr(maxclsize, rho[1])
+      Sigma2 <- corr(maxclsize, rho[2])
+      y1 <- c(t(rmvnorm(nclus, sigma = Sigma1)))
+      y2 <- c(t(rmvnorm(nclus, sigma = Sigma2)))
+      group <- rep(c(0, 1), each = nn)
+      if (!clusgrp) group  <- sample(group, nn, FALSE)
+      cid <- rep(1:(2 * nclus), each = maxclsize)
+      x <- exp(c(y1, y2)) + delta * group
+      dat <- data.frame(x = x, grp = group, cid = cid)
+      drop <-  sort(sample(1:(2 * nn), size = misrate * (2 * nn), FALSE))
+      if (misrate == 0.) dat else dat[-drop, ]
+  }
\end{Sinput}
\end{Schunk}
There are two required inputs: \code{nclus} for the number of clusters
in each group and \code{maxclsize} for the maximum cluster size.
The difference between groups is specified by \code{delta}.
The group specific intra-cluster correlation parameter $\rho_g$ is set
by \code{rho}, a numeric vector of length 2, one for each group.
The \code{corr} argument specifies the function to construct correlation
matrix with correlation coefficients in \code{rho}.
To allow unequal cluster sizes, \code{misrate} specifies the missing rate
at which an subunit in a cluster to be excluded at random; when
\code{misrate = 0}, balanced data with equal cluster size are generated.
The grouping level is controlled by the logical argument \code{clusgrp}:
\code{TRUE} for cluster-level and \code{FALSE} for subunit-level.
The function returns a data set with three columns: observation
\code{x}, grouping id \code{grp}, and cluster id \code{cid}.

For the replicability of the following demonstration,
a random seed is set.
\begin{Schunk}
\begin{Sinput}
R> set.seed(1234)
\end{Sinput}
\end{Schunk}

To illustrate the clustered rank-sum test, a data set with $10$
clusters of size $3$ in each group is generated with $\delta = 0$,
exchangeable correlation structure, $\rho_0 = \rho_1 = 0.9$,
and cluster-level treatment assignment.
\begin{Schunk}
\begin{Sinput}
R> dat.cl <- datgen.sum(10, 3, 0, c(.9, .9), ex, 0, TRUE)
\end{Sinput}
\end{Schunk}
The first and last 6 rows of the data frame looks like this:
\begin{Schunk}
\begin{Sinput}
R> cbind(head(dat.cl, 6), "head / tail" = "     ", tail(dat.cl, 6))
\end{Sinput}
\begin{Soutput}
          x grp cid head / tail         x grp cid
1 0.7322161   0   1             2.3997628   1  19
2 1.1708839   0   1             3.0184715   1  19
3 1.5112823   0   1             4.2536708   1  19
4 0.2516183   0   2             0.6759148   1  20
5 0.6050972   0   2             1.4343341   1  20
6 0.6199984   0   2             0.5985389   1  20
\end{Soutput}
\end{Schunk}

We first use the formula interface to perform the RGL asymptotic test:
\begin{Schunk}
\begin{Sinput}
R> clusWilcox.test(x ~ grp + cluster(cid), dat.cl, method = "rgl")
\end{Sinput}
\begin{Soutput}

	Clustered Wilcoxon rank sum test using Rosner-Glynn-Lee
	method

data:  x; group: grp; cluster: cid; (from dat.cl)
number of observations: 60;  number of clusters: 20
Z = -1.3613, p-value = 0.1734
alternative hypothesis: true difference in locations is not equal to 0
\end{Soutput}
\end{Schunk}
Setting \code{method = "ds"} would perform the DS test.
The exact test of the RGL method can be done for data with a
small number of clusters when treatment is assigned at cluster-level:
\begin{Schunk}
\begin{Sinput}
R> clusWilcox.test(x ~ grp + cluster(cid), dat.cl, method = "rgl",
+    exact = TRUE, B = 0)
\end{Sinput}
\begin{Soutput}

	Clustered Wilcoxon rank sum test using Rosner-Glynn-Lee
	method (exact exactutation)

data:  x; group: grp; cluster: cid; (from dat.cl)
number of observations: 60;  number of clusters: 20
W = 757, p-value = 0.1789
alternative hypothesis: true location is not equal to 0
\end{Soutput}
\end{Schunk}

The numerical interface is illustrated with the DS method:
\begin{Schunk}
\begin{Sinput}
R> clusWilcox.test(x, group = grp, cluster = cid, data = dat.cl, method = "ds")
\end{Sinput}
\begin{Soutput}

	Clustered Wilcoxon rank sum test using Datta-Satten method

data:  x; group: grp; cluster: cid; (from dat.cl)
number of observations: 60;  number of clusters: 20
Z = 1.3967, p-value = 0.1625
alternative hypothesis: true difference in locations is not equal to 0
\end{Soutput}
\end{Schunk}

The test statistics from the two methods are very close to each other.
Note that, when using the code provided online by the authors
of \citet{Rosn:Glyn:Lee:inco:2003} and \citet{Datt:Satt:rank:2005},
statistics with different signs may occur.
This is a result of using the opposite group to calculating the statistic.
To get the matching results, one just needs to switch the group ids.

For data with cluster-level groups, the RGL method allows an extra
stratum variable to accommodate stratification in the data.
For illustration, we simply add an extra column \code{strat} to the
dataset and perform the RGL test:
\begin{Schunk}
\begin{Sinput}
R> dat.cl$strat <- rep(rep(1:2, each = 15), 2)
R> cbind(head(dat.cl, 6), "head / tail" = "     ", tail(dat.cl, 6))
\end{Sinput}
\begin{Soutput}
          x grp cid strat head / tail         x grp cid strat
1 0.7322161   0   1     1             2.3997628   1  19     2
2 1.1708839   0   1     1             3.0184715   1  19     2
3 1.5112823   0   1     1             4.2536708   1  19     2
4 0.2516183   0   2     1             0.6759148   1  20     2
5 0.6050972   0   2     1             1.4343341   1  20     2
6 0.6199984   0   2     1             0.5985389   1  20     2
\end{Soutput}
\begin{Sinput}
R> clusWilcox.test(x ~ grp + cluster(cid) + stratum(strat), dat = dat.cl,
+    method = "rgl")
\end{Sinput}
\begin{Soutput}

	Clustered Wilcoxon rank sum test using Rosner-Glynn-Lee
	method

data:  x; group: grp; cluster: cid; stratum: strat; (from dat.cl)
number of observations: 60;  number of clusters: 20
Z = -1.3271, p-value = 0.1845
alternative hypothesis: true difference in locations is not equal to 0
\end{Soutput}
\end{Schunk}

The DS method can compare more than two groups.
We illustrate this by assigning 4 groups to this data:
\begin{Schunk}
\begin{Sinput}
R> dat.cl$grp <- rep(1:4, each = 15)
R> cbind(head(dat.cl, 6), "head / tail" = "     ", tail(dat.cl, 6))
\end{Sinput}
\begin{Soutput}
          x grp cid strat head / tail         x grp cid strat
1 0.7322161   1   1     1             2.3997628   4  19     2
2 1.1708839   1   1     1             3.0184715   4  19     2
3 1.5112823   1   1     1             4.2536708   4  19     2
4 0.2516183   1   2     1             0.6759148   4  20     2
5 0.6050972   1   2     1             1.4343341   4  20     2
6 0.6199984   1   2     1             0.5985389   4  20     2
\end{Soutput}
\begin{Sinput}
R> clusWilcox.test(x ~ grp + cluster(cid), dat = dat.cl, method = "ds")
\end{Sinput}
\begin{Soutput}

	Clustered Wilcoxon rank sum test using Datta-Satten method
	using Chi-square test

data:  x; group: grp; cluster: cid; (from dat.cl)
number of observations: 60;  number of clusters: 20
number of groups: 4
chi-square test statistic = 2.0471, p-value = 0.5627
\end{Soutput}
\end{Schunk}

\subsection{Signed-rank test for clustered data}

To illustrate clustered signed-rank tests, we generate data from a scheme
slightly modified from simulation scenario~1 in \citet{Datt:Satt:sign:2008}.
The paired differences in a cluster are generated as
\[
  X = \mathrm{sign}(Z) \exp(|Z|),
\]
where $Z$ is a multivariate normal random vector with mean $\delta$
and an exchangeable or AR1 correlation structure with parameter $\rho$.
This is implemented by the following function:
\begin{Schunk}
\begin{Sinput}
R> datgen.sgn <- function(nclus, maxclsize, delta = 0., rho = 0.1,
+    corr = ex, misrate = 0.) {
+      nn <- nclus * maxclsize
+      Sigma <- corr(maxclsize, rho)
+      z <- delta + c(t(rmvnorm(nclus, sigma = Sigma)))
+      x <- sign(z) * exp(abs(z))
+      cid <- rep(1:nclus, each = maxclsize)
+      dat <- data.frame(x = x, cid = cid)
+      drop <- sort(sample(1:nn, size = misrate * nn, FALSE))
+      if (misrate == 0.) dat else dat[-drop,]
+  }
\end{Sinput}
\end{Schunk}
The arguments of \code{datgen.sgn} match those of \code{datgen.sum},
except that it does not need a \code{clusgrp} argument because the
data are already differences between two groups.

For illustration, we generate a dataset that consists of $10$ clusters of
size $3$, with exchangeable correlation parameter $\rho = 0.5$.
\begin{Schunk}
\begin{Sinput}
R> dat.sgn <- datgen.sgn(10, 3, cor = ex, rho = 0.5)
R> cbind(head(dat.cl, 6), "head / tail" = "     ", tail(dat.cl, 6))
\end{Sinput}
\begin{Soutput}
          x grp cid strat head / tail         x grp cid strat
1 0.7322161   1   1     1             2.3997628   4  19     2
2 1.1708839   1   1     1             3.0184715   4  19     2
3 1.5112823   1   1     1             4.2536708   4  19     2
4 0.2516183   1   2     1             0.6759148   4  20     2
5 0.6050972   1   2     1             1.4343341   4  20     2
6 0.6199984   1   2     1             0.5985389   4  20     2
\end{Soutput}
\end{Schunk}

The RGL signed-rank test is performed with:
\begin{Schunk}
\begin{Sinput}
R> clusWilcox.test(x ~ cluster(cid),  dat.sgn, paired = TRUE, method = "rgl")
\end{Sinput}
\begin{Soutput}

	Clustered Wilcoxon signed rank test using Rosner-Glynn-Lee
	method

data:  x; cluster: cid; (from dat.sgn)
number of observations: 30;  number of clusters: 10
Z = 0.47709, p-value = 0.6333
alternative hypothesis: true shift in location is not equal to 0
\end{Soutput}
\end{Schunk}
The DS signed-rank test is performed with
\begin{Schunk}
\begin{Sinput}
R> clusWilcox.test(x ~ cluster(cid),  dat.sgn, paired = TRUE, method = "ds")
\end{Sinput}
\begin{Soutput}

	Clustered Wilcoxon signed rank test using Datta-Satten method

data:  x; cluster: cid; (from dat.sgn)
number of observations: 30;  number of clusters: 10
Z = 0.45109, p-value = 0.6519
alternative hypothesis: true shift in location is not equal to 0
\end{Soutput}
\end{Schunk}

\subsection{A real data example}

The package contains a real dataset named \code{amd} from a retrospective
observational study \citep{Ferr:Sedd:phen:2015} that aims to characterize the
phenotype associated with a rare variant in the complement factor H (CFH)
R1210C,  a protein involved in the age-related macular degeneration (AMD).
The data contains measures of 283 eyes from 143 patients,
among whom 62 had the rare variant and 81 did not.
The clusters are the patients and the subunits are the eyes.
The outcome variable was the AMD grading score based
on the Age-Related Maculopathy Staging (CARMS),  which has 5~levels.
The first 3~levels correspond to the size of drusen which is an
intermediate marker of AMD, while level~4 and~5 correspond to different
types of AMD, with level~4 indicating the presence
of geographic atrophy (GA) and level~5 indicating the presence of
choroidal neovascularization (CNV).
The Wilcoxon rank-sum test is to be carried out on two subsets:
1) the subset of observations with CARMS grade 1, 2, 3 or~4;
and 2) the subset of observations with CARMS grade 1, 2, 3 or~5.
In the \code{amd} data, the outcome variable is \code{CARMS};
the patients ids indicating the clusters are stored in variable \code{ID};
the rare variant grouping at the cluster (patient) level is indicated by
variable \code{Variant}.

We apply both the RGL and DS method to the first subset:
\begin{Schunk}
\begin{Sinput}
R> data(amd)
R> clusWilcox.test(CARMS ~ Variant + cluster(ID), data = amd,
+    subset = CARMS %in% c(1, 2, 3, 4), method = "rgl")
\end{Sinput}
\begin{Soutput}

	Clustered Wilcoxon rank sum test using Rosner-Glynn-Lee
	method

data:  CARMS; group: Variant; cluster: ID; (from amd)
number of observations: 196;  number of clusters: 112
Z = -4.3993, p-value = 1.086e-05
alternative hypothesis: true difference in locations is not equal to 0
\end{Soutput}
\begin{Sinput}
R> clusWilcox.test(CARMS ~ Variant + cluster(ID), data = amd,
+    subset = CARMS %in% c(1, 2, 3, 4), method = "ds")
\end{Sinput}
\begin{Soutput}

	Clustered Wilcoxon rank sum test using Datta-Satten method

data:  CARMS; group: Variant; cluster: ID; (from amd)
number of observations: 196;  number of clusters: 112
Z = -4.4823, p-value = 7.384e-06
alternative hypothesis: true difference in locations is not equal to 0
\end{Soutput}
\end{Schunk}
The p-values from both tests are close and less than $0.001$, which
implies strong evidence of an association between the presence of
CFH~R1210C rare variant and the CARMS grade with GA as the advanced stage.

For the RGL method, a stratifying variable \code{Agesex} which categorizes
the patients into 6~strata by their age and gender can be used as a control
variable:
\begin{Schunk}
\begin{Sinput}
R> clusWilcox.test(CARMS ~ Variant + cluster(ID) + stratum(Agesex), data = amd,
+    subset = CARMS %in% c(1, 2, 3, 4))
\end{Sinput}
\begin{Soutput}

	Clustered Wilcoxon rank sum test using Rosner-Glynn-Lee
	method

data:  CARMS; group: Variant; cluster: ID; stratum: Agesex; (from amd)
number of observations: 196;  number of clusters: 112
Z = -4.0797, p-value = 4.509e-05
alternative hypothesis: true difference in locations is not equal to 0
\end{Soutput}
\end{Schunk}
The p-value is still less than $0.001$ after controlling for age and gender.

We then apply the two tests to the second subset:
\begin{Schunk}
\begin{Sinput}
R> clusWilcox.test(CARMS ~ Variant + cluster(ID), data = amd, method = "rgl",
+    subset = CARMS %in% c(1, 2, 3, 5))
\end{Sinput}
\begin{Soutput}

	Clustered Wilcoxon rank sum test using Rosner-Glynn-Lee
	method

data:  CARMS; group: Variant; cluster: ID; (from amd)
number of observations: 224;  number of clusters: 121
Z = -1.8484, p-value = 0.06455
alternative hypothesis: true difference in locations is not equal to 0
\end{Soutput}
\begin{Sinput}
R> clusWilcox.test(CARMS ~ Variant + cluster(ID), data = amd, method = "ds",
+    subset = CARMS %in% c(1, 2, 3, 5))
\end{Sinput}
\begin{Soutput}

	Clustered Wilcoxon rank sum test using Datta-Satten method

data:  CARMS; group: Variant; cluster: ID; (from amd)
number of observations: 224;  number of clusters: 121
Z = -2.7311, p-value = 0.006312
alternative hypothesis: true difference in locations is not equal to 0
\end{Soutput}
\end{Schunk}
This time, the p-values of the two approaches are easily discernable, which
is quite possible because the methods are based on different assumptions.
The DS method reports a p-value of $0.006312$, in contrast to $0.06455$ from
the RGL method. The results suggest association between the presence of
CFH~R1210C rare variant and the symptom with CNV as the advanced stage.
Again, the RGL method can be applied with age and gender controlled:
\begin{Schunk}
\begin{Sinput}
R> clusWilcox.test(CARMS ~ Variant + cluster(ID) + stratum(Agesex), data = amd,
+    subset = CARMS %in% c(1, 2, 3, 5), method = "rgl")
\end{Sinput}
\begin{Soutput}

	Clustered Wilcoxon rank sum test using Rosner-Glynn-Lee
	method

data:  CARMS; group: Variant; cluster: ID; stratum: Agesex; (from amd)
number of observations: 224;  number of clusters: 121
Z = -1.8519, p-value = 0.06404
alternative hypothesis: true difference in locations is not equal to 0
\end{Soutput}
\end{Schunk}
The p-value from the RGL method remains virtually unchanged after
controlling for the age/gender strata.

\section{A simulation study}
\label{s:simu}

Comparisons of the RGL and DS methods under some common
scenarios have not been studied in the recent literature.
Such comparison can be easily done with the package \pkg{clusrank}.
Using the two data generation functions defined above, we conduct a
simulation study comparing their sizes and powers in a few settings.
The following function generates replicates of data for a
given scenario and returns the empirical power for a given
significance level.
\begin{Schunk}
\begin{Sinput}
R> simpower <- function(nrep, level, paired, nclus, maxclsize,
+    delta, rho, corr, misrate, ...) {
+      do1rep <- function() {
+        datgen <- if (paired) datgen.sgn else datgen.sum
+        formula <- if (paired) x ~ cluster(cid)
+                   else x ~ cluster(cid) + grp
+        dat <- datgen(nclus, maxclsize, delta, rho, corr, misrate, ...)
+        p.rgl <- clusWilcox.test(formula, paired = paired,
+                                 data = dat, method = "rgl")$p.value
+        p.ds  <- clusWilcox.test(formula, paired = paired,
+                                 data = dat, method = "ds" )$p.value
+        c(rgl = p.rgl, ds = p.ds)
+      }
+      sim <- t(replicate(nrep, do1rep()))
+      apply(sim, 2, function(x) mean(x < level))
+  }
\end{Sinput}
\end{Schunk}
The first two arguments, \code{nrep} and \code{level} specify the
number of replications and the desired significance level, respectively.
Argument \code{paired} is a logical scalar to switch between the two methods,
\code{TRUE} for signed-rank tests and \code{FALSE} for rank-sum tests.
Other arguments have the same meanings as those in \code{datgen.sum}
or \code{datgen.sgn} depending on the value of \code{paired}.
The last argument \code{...} is used to supply \code{clusgrp}, which
is only needed by \code{datgen.sum} for the level of groups.
The function returns the empirical rejection rates of the RGL
and DS methods for the setting defined by the inputs.

As an example, consider the rank-sum tests in a setting with
cluster-level grouping, each group containing 20~clusters of size~3, with
exchangeable correlation parameter $\rho = 0.5$ in both groups.
The group difference is set to be $\delta = 0$.
We do this experiment with 1000 replicates at significance level 0.05:
\begin{Schunk}
\begin{Sinput}
R> simpower(1000, 0.05, FALSE, 20, 3, 0.0, c(0.5, 0.5), ex, 0., clusgrp = TRUE)
\end{Sinput}
\begin{Soutput}
  rgl    ds 
0.052 0.056 
\end{Soutput}
\end{Schunk}
The empirical sizes of both methods are close to the nominal level $0.05$.

Similarly, a comparison for signed-rank tests can be done.
This time we use an AR1 correlation setting with $\rho = 0.5$.
\begin{Schunk}
\begin{Sinput}
R> simpower(1000, 0.05, TRUE,  20, 3, 0.0, 0.5, ar1, 0.)
\end{Sinput}
\begin{Soutput}
  rgl    ds 
0.055 0.057 
\end{Soutput}
\end{Schunk}
Again, both methods have empirical sizes close to the nominal level.
It means that the RGL method is robust to the violation of the
exchangeability assumption in this setting.

\subsection{Rank-sum tests for clustered data}

\begin{table}[tbp]
\centering
\begin{tabular}{ccccc rr rr rr}
  \toprule
Group & Missing & Max & Corr & Group &\multicolumn{2}{c}{$\delta = 0$} & \multicolumn{2}{c}{$\delta = 0.2$} & \multicolumn{2}{c}{$\delta = 0.5$}  \\
    \cmidrule(lr){6-7} \cmidrule(lr){8-9} \cmidrule(lr){10-11}
 level & rate & $n_i$ & $\rho$ & size  & RGL & DS & RGL & DS & RGL & DS \\
  \midrule
cluster & 0 & 2 & 0.1, 0.1 & 20 & 4.3 & 4.5 & 16.3 & 17.0 & 64.1 & 65.2 \\
   &  &  &  & 50 & 4.7 & 4.8 & 36.5 & 37.0 & 96.4 & 96.5 \\
   &  &  & 0.5, 0.5 & 20 & 5.2 & 5.5 & 14.7 & 15.1 & 52.5 & 53.7 \\
   &  &  &  & 50 & 5.0 & 5.0 & 29.8 & 29.9 & 89.6 & 89.6 \\
   &  &  & $-$0.1, 0.9 & 20 & 5.3 & 5.5 & 14.6 & 15.3 & 59.2 & 60.2 \\
   &  &  &  & 50 & 4.8 & 4.9 & 30.6 & 30.9 & 95.0 & 95.1 \\
   &  & 5 & 0.1, 0.1 & 20 & 4.8 & 5.0 & 28.9 & 30.0 & 91.7 & 92.1 \\
   &  &  &  & 50 & 5.0 & 5.1 & 64.1 & 64.4 & 99.9 & 99.9 \\
   &  &  & 0.5, 0.5 & 20 & 5.0 & 5.5 & 16.6 & 17.4 & 64.0 & 64.8 \\
   &  &  &  & 50 & 4.7 & 4.8 & 36.6 & 37.1 & 95.6 & 95.7 \\
   &  &  & $-$0.1, 0.9 & 20 & 5.3 & 5.5 & 18.2 & 18.9 & 79.7 & 80.6 \\
   &  &  &  & 50 & 4.8 & 4.8 & 42.0 & 42.4 & 99.6 & 99.7 \\
   & 0.5 & 10 & 0.1, 0.1 & 20 & 5.7 & 6.0 & 24.6 & 28.2 & 84.3 & 89.5 \\
   &  &  &  & 50 & 4.7 & 4.8 & 58.4 & 59.1 & 100.0 & 99.9 \\
   &  &  & 0.5, 0.5 & 20 & 4.7 & 5.0 & 13.6 & 16.9 & 53.0 & 62.6 \\
   &  &  &  & 50 & 4.7 & 5.2 & 29.7 & 33.4 & 92.4 & 94.9 \\
   &  &  & $-$0.1, 0.9 & 20 & 4.8 & 4.5 & 14.1 & 16.9 & 67.5 & 77.2 \\
   &  &  &  & 50 & 4.9 & 4.8 & 36.5 & 41.2 & 99.4 & 99.5 \\
    subunit & 0 & 2 & 0.1, 0.1 & 20 & 5.7 & 5.8 & 18.3 & 18.5 & 69.7 & 68.6 \\
   &  &  &  & 50 & 4.7 & 4.7 & 41.2 & 41.4 & 97.6 & 97.5 \\
   &  &  & 0.5, 0.5 & 20 & 4.5 & 4.5 & 19.0 & 18.9 & 68.7 & 67.3 \\
   &  &  &  & 50 & 4.7 & 4.8 & 41.5 & 41.2 & 97.5 & 97.2 \\
   &  &  & $-$0.1, 0.9 & 20 & 4.6 & 4.9 & 19.6 & 19.2 & 70.5 & 68.2 \\
   &  &  &  & 50 & 4.6 & 4.7 & 41.3 & 40.5 & 97.5 & 97.1 \\
   &  & 5 & 0.1, 0.1 & 20 & 4.7 & 4.6 & 40.2 & 38.6 & 97.5 & 96.8 \\
   &  &  &  & 50 & 4.9 & 5.0 & 77.2 & 76.9 & 100.0 & 100.0 \\
   &  &  & 0.5, 0.5 & 20 & 5.0 & 5.0 & 43.3 & 40.5 & 97.1 & 95.7 \\
   &  &  &  & 50 & 5.3 & 5.5 & 77.4 & 75.7 & 100.0 & 100.0 \\
   &  &  & $-$0.1, 0.9 & 20 & 5.2 & 4.8 & 43.4 & 40.1 & 96.6 & 95.0 \\
   &  &  &  & 50 & 5.0 & 5.1 & 78.4 & 76.5 & 100.0 & 100.0 \\
   & 0.5 & 10 & 0.1, 0.1 & 20 & 4.9 & 5.0 & 38.3 & 35.8 & 96.5 & 94.5 \\
   &  &  &  & 50 & 4.7 & 4.3 & 76.4 & 71.5 & 100.0 & 100.0 \\
   &  &  & 0.5, 0.5 & 20 & 4.5 & 4.7 & 42.8 & 35.2 & 97.1 & 93.1 \\
   &  &  &  & 50 & 3.6 & 4.4 & 78.2 & 70.3 & 100.0 & 100.0 \\
   &  &  & $-$0.1, 0.9 & 20 & 4.5 & 4.8 & 44.9 & 34.8 & 97.4 & 92.6 \\
   &  &  &  & 50 & 4.8 & 4.7 & 79.8 & 70.4 & 100.0 & 99.9 \\
   \bottomrule
\end{tabular}
\caption{Empirical rejection percentage of the RGL and the DS methods
  for rank-sum tests at nominal significance level 0.05 when
  intracluster correlation is exchangable.
  The results are based on 4000 datasets. Each group contains
  same number of clusters.}
\label{tab:rsex}
\end{table}

\begin{table}[tbp]
\centering
\begin{tabular}{ccccc rr rr rr}
  \toprule
Group & Missing & Max & Corr & Group &\multicolumn{2}{c}{$\delta = 0$} & \multicolumn{2}{c}{$\delta = 0.2$} & \multicolumn{2}{c}{$\delta = 0.5$}  \\
    \cmidrule(lr){6-7} \cmidrule(lr){8-9} \cmidrule(lr){10-11}
 level & rate & $n_i$ & $\rho$ & size  & RGL & DS & RGL & DS & RGL & DS \\
  \midrule
cluster & 0 & 2 & 0.1, 0.1 & 20 & 5.0 & 5.3 & 19.3 & 20.0 & 63.8 & 64.8 \\
   &  &  &  & 50 & 4.8 & 4.9 & 35.9 & 36.5 & 96.2 & 96.2 \\
   &  &  & 0.5, 0.5 & 20 & 5.6 & 5.9 & 14.0 & 14.5 & 52.6 & 53.9 \\
   &  &  &  & 50 & 5.2 & 5.3 & 28.1 & 28.3 & 90.1 & 90.2 \\
   &  &  & $-$0.1, 0.9 & 20 & 4.5 & 4.9 & 13.7 & 14.4 & 58.9 & 60.3 \\
   &  &  &  & 50 & 4.6 & 4.8 & 31.9 & 32.1 & 94.2 & 94.5 \\
   &  & 5 & 0.1, 0.1 & 20 & 5.0 & 5.3 & 34.3 & 35.4 & 93.6 & 93.8 \\
   &  &  &  & 50 & 4.5 & 4.7 & 69.5 & 69.9 & 100.0 & 100.0 \\
   &  &  & 0.5, 0.5 & 20 & 4.3 & 4.9 & 20.3 & 21.1 & 75.0 & 75.9 \\
   &  &  &  & 50 & 5.1 & 5.2 & 46.4 & 46.9 & 98.9 & 98.9 \\
   &  &  & $-$0.1, 0.9 & 20 & 5.1 & 5.3 & 19.3 & 20.0 & 78.6 & 79.5 \\
   &  &  &  & 50 & 5.1 & 5.2 & 42.2 & 42.7 & 99.6 & 99.6 \\
   & 0.5 & 10 & 0.1, 0.1 & 20 & 4.8 & 5.0 & 30.6 & 34.4 & 91.6 & 93.6 \\
   &  &  &  & 50 & 4.8 & 5.0 & 70.7 & 68.7 & 100.0 & 100.0 \\
   &  &  & 0.5, 0.5 & 20 & 4.2 & 4.8 & 20.6 & 24.1 & 76.2 & 83.7 \\
   &  &  &  & 50 & 5.8 & 5.6 & 49.2 & 52.5 & 99.7 & 99.8 \\
   &  &  & $-$0.1, 0.9 & 20 & 5.3 & 5.2 & 15.9 & 19.3 & 70.0 & 79.3 \\
   &  &  &  & 50 & 5.0 & 5.1 & 40.0 & 43.7 & 99.2 & 99.6 \\
  subunit & 0 & 2 & 0.1, 0.1 & 20 & 4.8 & 5.3 & 19.3 & 19.3 & 69.3 & 68.0 \\
   &  &  &  & 50 & 5.0 & 5.0 & 40.4 & 40.4 & 97.2 & 96.9 \\
   &  &  & 0.5, 0.5 & 20 & 4.8 & 4.9 & 19.5 & 18.9 & 69.1 & 67.5 \\
   &  &  &  & 50 & 5.0 & 4.9 & 40.9 & 40.1 & 98.1 & 97.7 \\
   &  &  & $-$0.1, 0.9 & 20 & 4.4 & 4.5 & 19.1 & 18.3 & 70.0 & 67.8 \\
   &  &  &  & 50 & 5.0 & 5.0 & 40.4 & 39.1 & 97.1 & 96.9 \\
   &  & 5 & 0.1, 0.1 & 20 & 5.1 & 4.9 & 39.5 & 37.8 & 97.6 & 97.0 \\
   &  &  &  & 50 & 4.8 & 4.8 & 77.5 & 77.1 & 100.0 & 100.0 \\
   &  &  & 0.5, 0.5 & 20 & 4.8 & 4.9 & 42.3 & 38.6 & 97.0 & 95.8 \\
   &  &  &  & 50 & 4.7 & 4.7 & 78.1 & 76.9 & 100.0 & 100.0 \\
   &  &  & $-$0.1, 0.9 & 20 & 5.0 & 4.6 & 41.1 & 37.0 & 97.2 & 95.8 \\
   &  &  &  & 50 & 4.7 & 4.8 & 77.8 & 76.0 & 100.0 & 100.0 \\
   & 0.5 & 10 & 0.1, 0.1 & 20 & 5.2 & 5.5 & 38.5 & 35.8 & 96.5 & 94.3 \\
   &  &  &  & 50 & 5.5 & 4.9 & 76.5 & 69.8 & 100.0 & 100.0 \\
   &  &  & 0.5, 0.5 & 20 & 5.2 & 5.3 & 43.2 & 34.8 & 97.2 & 93.5 \\
   &  &  &  & 50 & 5.3 & 5.6 & 78.8 & 71.3 & 100.0 & 100.0 \\
   &  &  & $-$0.1, 0.9 & 20 & 4.8 & 4.0 & 46.5 & 35.4 & 97.7 & 93.0 \\
   &  &  &  & 50 & 5.5 & 5.4 & 81.6 & 72.0 & 100.0 & 100.0 \\
   \bottomrule
\end{tabular}
\caption{Empirical rejection percentage of the RGL and the DS methods
  for rank-sum tests at nominal significance level 0.05 when
  intracluster correlation is AR1. The results are based on
  4000 datasets. Each group contains same number of clusters.}
\label{tab:rsar}
\end{table}

Comparison between the RGL and DS methods for the rank-sum tests
has never been done previously under subunit-level treatment group.
Nor has it been done when the intracluster dependence is not exchangeable.
Therefore we will present these comparisons in this session. Consider
two equal size groups, with 20 or 50~clusters, and with exchangeable
or AR1 intracluster correlation structure as specified in \code{datgen.sum}.
The correlation parameters for the two groups were set to be
$(0.1, 0.1)$, $(0.5, 0.5)$, or $(-0.1, 0.9)$.
The maximum cluster size $G = \max_i n_i$ has three levels: 2, 5, and 10.
The \code{simpower} function makes the comparison very easy, and the
\code{misrate} argument allows cluster sizes to be random, which is
an additional scenario of interest that has not been compared before.
Two missing rates, $0$ and $0.5$, were considered, representing
balanced data and unbalanced data with missing completely at random, respectively.
This is different from the informative cluster size setting studied
in \citet{Datt:Satt:rank:2005}, where cluster size depends on group.
The group difference $\delta$ was set to be $0$, $0.2$ and $0.5$.
For each setting, empirical rejection rates was obtained from 4000 replicates.

The results are summarized in Table~\ref{tab:rsex}--\ref{tab:rsar}.
The empirical size ($\delta = 0$) of both methods are close to the
nominal level $0.05$ in all the settings considered in this study.
The empirical power ($\delta \ne 0$) for both methods increases as the
cluster size increases, and as the missing rate decreases.
For balanced data, the powers of the two tests are very close regardless
of the level of the treatment group assignment and the intracluster configurations.
For unbalanced data from maximum cluster size~10 and missing rate 0.5,
the DS method has higher power with cluster-level group assignment, while
the RGL method has higher power with subunit-level group assignment.
As this setting has average cluster size~5, it can be compared with
the balanced data setting with cluster size fixed at~5.
Both methods have higher power in the random cluster size setting
when the treatment group is assigned at the subunit-level.
Under cluster-level group assignment, it appears that the RGL method has
lower power with random cluster size, while the DS method has lower
power with fixed cluster size, though the differences are not big.
Both tests seem to be robust to the intracluster correlation structure.

\subsection{Signed-rank tests for clustered data}

\begin{table}[tbp]
  \centering
\begin{tabular}{cccc  rr rr rr}
  \toprule
 Missing & Max & Corr & Group & \multicolumn{2}{c}{$\delta = 0$} & \multicolumn{2}{c}{$\delta = 0.2$} & \multicolumn{2}{c}{$\delta = 0.5$} \\
  \cmidrule(lr){5-6} \cmidrule(lr){7-8} \cmidrule(lr){9-10}
  rate & $n_i$ & $\rho$ & size  & RGL & DS & RGL & DS & RGL & DS \\
  \midrule
 0 & 2 & 0.1 & 20 & 5.4 & 5.6 & 19.7 & 20.4 & 78.8 & 79.0 \\
   &  &  & 50 & 5.0 & 5.1 & 44.2 & 44.4 & 99.6 & 99.6 \\
   &  & 0.5 & 20 & 5.0 & 5.3 & 15.4 & 15.8 & 68.1 & 69.0 \\
   &  &  & 50 & 5.1 & 5.2 & 34.6 & 34.9 & 97.6 & 97.7 \\
   &  & 0.9 & 20 & 5.1 & 5.4 & 13.7 & 14.1 & 56.5 & 57.4 \\
   &  &  & 50 & 4.2 & 4.2 & 29.4 & 29.8 & 93.7 & 93.8 \\
   & 10 & 0.1 & 20 & 4.4 & 4.5 & 47.2 & 47.5 & 99.6 & 99.6 \\
   &  &  & 50 & 4.6 & 4.7 & 88.0 & 88.0 & 100.0 & 100.0 \\
   &  & 0.5 & 20 & 5.0 & 5.2 & 19.7 & 20.2 & 81.4 & 81.9 \\
   &  &  & 50 & 4.8 & 4.9 & 45.2 & 45.5 & 99.6 & 99.6 \\
   &  & 0.9 & 20 & 5.0 & 5.3 & 13.6 & 14.3 & 58.9 & 59.9 \\
   &  &  & 50 & 5.1 & 5.1 & 29.4 & 30.0 & 95.0 & 95.1 \\
  0.5 & 5 & 0.1 & 20 & 4.4 & 4.8 & 21.4 & 20.2 & 82.7 & 78.3 \\
   &  &  & 50 & 5.2 & 5.0 & 50.6 & 45.3 & 99.8 & 99.5 \\
   &  & 0.5 & 20 & 4.6 & 5.2 & 15.6 & 15.7 & 67.2 & 67.1 \\
   &  &  & 50 & 4.7 & 4.8 & 36.0 & 35.2 & 97.9 & 97.5 \\
   &  & 0.9 & 20 & 4.9 & 5.2 & 14.3 & 14.8 & 58.1 & 59.3 \\
   &  &  & 50 & 4.5 & 4.7 & 27.9 & 28.1 & 93.6 & 93.7 \\
   & 10 & 0.1 & 20 & 5.2 & 5.1 & 33.1 & 31.9 & 96.9 & 96.5 \\
   &  &  & 50 & 5.1 & 5.0 & 72.3 & 69.7 & 100.0 & 100.0 \\
   &  & 0.5 & 20 & 4.3 & 4.5 & 17.7 & 18.0 & 77.7 & 77.7 \\
   &  &  & 50 & 5.1 & 5.2 & 41.3 & 41.1 & 99.3 & 99.3 \\
   &  & 0.9 & 20 & 4.5 & 4.8 & 13.1 & 13.7 & 58.1 & 59.2 \\
   &  &  & 50 & 5.1 & 5.2 & 30.0 & 30.2 & 93.9 & 93.9 \\
  \bottomrule
\end{tabular}
   \caption{
     Empirical rejection percentage of the RGL and DS methods for
     signed-rank tests at nominal significance level 0.05 with
     exchangable intracluster correlation. The results are based on
     4000 datasets. Each group contains same number of clusters.}
\label{tab:srex}
\end{table}

\begin{table}[tbp]
  \centering
\begin{tabular}{cccc  rr rr rr}
  \toprule
 Missing & Max & Corr & Group & \multicolumn{2}{c}{$\delta = 0$} & \multicolumn{2}{c}{$\delta = 0.2$} & \multicolumn{2}{c}{$\delta = 0.5$} \\
  \cmidrule(lr){5-6} \cmidrule(lr){7-8} \cmidrule(lr){9-10}
  rate & $n_i$ & $\rho$ & size & RGL & DS & RGL & DS & RGL & DS \\
  \midrule
  0 & 2 & 0.1 & 20 & 4.35 & 4.60 & 21.12 & 21.25 & 78.22 & 78.60 \\
   &  &  & 50 & 5.00 & 5.10 & 44.50 & 44.75 & 99.47 & 99.47 \\
   &  & 0.5 & 20 & 4.25 & 4.42 & 15.38 & 15.85 & 68.03 & 68.80 \\
   &  &  & 50 & 5.35 & 5.40 & 35.70 & 36.10 & 97.45 & 97.53 \\
   &  & 0.9 & 20 & 5.10 & 5.40 & 13.38 & 13.90 & 56.58 & 57.33 \\
   &  &  & 50 & 4.08 & 4.20 & 29.70 & 30.05 & 93.62 & 93.90 \\
   & 10 & 0.1 & 20 & 4.17 & 4.33 & 65.37 & 65.05 & 100.00 & 100.00 \\
   &  &  & 50 & 4.60 & 4.58 & 97.65 & 97.53 & 100.00 & 100.00 \\
   &  & 0.5 & 20 & 4.90 & 5.03 & 36.23 & 36.73 & 98.47 & 98.53 \\
   &  &  & 50 & 4.60 & 4.60 & 77.10 & 77.17 & 100.00 & 100.00 \\
   &  & 0.9 & 20 & 4.72 & 4.95 & 15.75 & 16.25 & 68.92 & 69.78 \\
   &  &  & 50 & 4.62 & 4.70 & 35.85 & 35.98 & 98.03 & 98.10 \\
  0.5 & 5 & 0.1 & 20 & 4.10 & 4.55 & 22.48 & 20.30 & 87.40 & 82.22 \\
   &  &  & 50 & 5.17 & 5.40 & 55.65 & 49.65 & 99.92 & 99.72 \\
   &  & 0.5 & 20 & 5.50 & 5.58 & 17.15 & 17.20 & 75.60 & 73.72 \\
   &  &  & 50 & 4.88 & 4.85 & 41.40 & 39.88 & 98.80 & 98.47 \\
   &  & 0.9 & 20 & 5.15 & 5.67 & 13.72 & 14.30 & 59.70 & 60.40 \\
   &  &  & 50 & 4.67 & 4.95 & 28.50 & 28.93 & 93.95 & 93.92 \\
   & 10 & 0.1 & 20 & 4.75 & 4.67 & 40.77 & 37.70 & 99.15 & 98.28 \\
   &  &  & 50 & 5.00 & 4.90 & 82.33 & 78.75 & 100.00 & 100.00 \\
   &  & 0.5 & 20 & 4.95 & 5.17 & 29.07 & 28.60 & 93.65 & 93.05 \\
   &  &  & 50 & 4.97 & 4.78 & 62.55 & 61.65 & 100.00 & 100.00 \\
   &  & 0.9 & 20 & 4.75 & 4.97 & 16.40 & 17.25 & 67.03 & 67.08 \\
   &  &  & 50 & 4.60 & 4.70 & 34.50 & 34.38 & 97.40 & 97.42 \\
  \bottomrule
\end{tabular}
\caption{
     Empirical rejection percentage of the RGL and DS methods for
     signed-rank tests at nominal significance level 0.05 with
     AR1 intracluster correlation. The results are based on
     4000 datasets. Each group contains same number of clusters.}
\label{tab:srar}
\end{table}

For the signed-rank test, \citet{Datt:Satt:sign:2008} did not have
settings with completely random cluster size, and their non-exchangeable
intracluster dependence is different from our AR1 setting.
We considered settings similar to those for the rank-sum test.
The paired differences were generated for 20 or 50 clusters.
The intracluster correlation parameter was set to be 0.1, 0.5, and 0.9.
The true difference $\delta$ was set to be $0$, $0.2$ and $0.5$.
For each setting, 4000 replicates were generated.

Selected results are summarized in Table~\ref{tab:srex}--\ref{tab:srar}.
In all settings in this study, the empirical sizes of both methods are
close to the nominal level, including the cases under of AR1 correlation
where the exchangeability assumption for the RGL method is violated
The empirical power of both tests increases as the cluster size increases,
and as the intracluster dependence decreases.
Completely random cluster size reduces the powers in comparison to
the cases where the cluster sizes are fixed at their means.
In all the settings considered here, the two methods preformed similarly.

\section{Discussion}
\label{s:disc}

Clustered data are frequently encountered in scientific research, and
rank-based tests for clustered data are an indispensable tool like
their counterparts, the Wilcoxon tests for independent data.
The package \pkg{clusrank} provides two schools, the RGL method and the DS
method, of the recently developed rank-sum tests and signed-rank tests
in a unified, higher-level, user-friendly interface.
Users need to be aware of the applicability of these tests when using them.
For example, the RGL method assumes exchangeability within clusters
and does not account for informative cluster sizes;
the DS method cannot be applied to contralateral designs with exactly one
subunit in each group within a cluster; both asymptotic tests require that
the number of clusters to not to be too small.

Implementation of other rank-based methods for clustered data can be
considered in future development of the package \pkg{clusrank}.
For the rank-sum test, the case of informative group size within a cluster
\citep{Dutt:Datt:rank:2016} would be a useful addition, though it is
available in the package \pkg{ClusterRankTest} \citep{Dutt:Datt:clus:2016}.
For the signed-rank test,  \citet{Laro:wilc:2005} has a variance estimator
based on certain sums of squares over independent clusters.
Sign tests and signed-rank tests for multivariate clustered data
\citep{Laro:affi:2003, Laro:Neva:Oja:weig:2007, Haat:etal:weig:2009} and
multilevel data \citep{Laro:Neva:Oja:one:2008} have been studied.
Some tests allow the distributions in two groups to have different scales
and/or shapes under the null hypothesis \citep{Laro:etal:two:2010}.
Making these tests available would be of interest to many users.
When covariates are available, rank regression for clustered data
\citep[e.g.,][]{Wang:Zhu:rank:2006, Wang:Zhao:weig:2008,
  Fu:Wang:Bai:rank:2010} would be the next step.

\section*{Acknowledgement}

We thank Dr. Johanna Seddon for sharing the data from their
phenotypic characterization study of the H~R1210C rare variant.
Rosner and Lee were supported in part by grant NIH R01EY022445
from the National Eye Institute.

\bibliography{clusrank}

\end{document}